\documentclass[reprint,amsmath,amssymb,aps]{revtex4-1}

\usepackage{graphicx}
\usepackage{dcolumn}
\usepackage{bm}
\usepackage{hyperref}
\usepackage{amsmath}
\usepackage{amssymb}
\usepackage{xcolor}
\usepackage{ulem}

\begin{document}

\title{Efficient and Robust Entanglement Generation with Deep Reinforcement Learning for Quantum Metrology}

\author{Yuxiang Qiu}
\affiliation{Guangdong Provincial Key Laboratory of Quantum Metrology and Sensing $\&$ School of Physics and Astronomy, Sun Yat-Sen University (Zhuhai Campus), Zhuhai 519082, China}
\affiliation{State Key Laboratory of Optoelectronic Materials and Technologies, Sun Yat-Sen University (Guangzhou Campus), Guangzhou 510275, China}

\author{Min Zhuang}
\affiliation{Guangdong Provincial Key Laboratory of Quantum Metrology and Sensing $\&$ School of Physics and Astronomy, Sun Yat-Sen University (Zhuhai Campus), Zhuhai 519082, China}
\affiliation{State Key Laboratory of Optoelectronic Materials and Technologies, Sun Yat-Sen University (Guangzhou Campus), Guangzhou 510275, China}

\author{Jiahao Huang}
\altaffiliation{Email: hjiahao@mail2.sysu.edu.cn, eqjiahao@gmail.com}
\affiliation{Guangdong Provincial Key Laboratory of Quantum Metrology and Sensing $\&$ School of Physics and Astronomy, Sun Yat-Sen University (Zhuhai Campus), Zhuhai 519082, China}
\affiliation{State Key Laboratory of Optoelectronic Materials and Technologies, Sun Yat-Sen University (Guangzhou Campus), Guangzhou 510275, China}

\author{Chaohong Lee}
\altaffiliation{Email: lichaoh2@mail.sysu.edu.cn, chleecn@gmail.com}
\affiliation{Guangdong Provincial Key Laboratory of Quantum Metrology and Sensing $\&$ School of Physics and Astronomy, Sun Yat-Sen University (Zhuhai Campus), Zhuhai 519082, China}
\affiliation{State Key Laboratory of Optoelectronic Materials and Technologies, Sun Yat-Sen University (Guangzhou Campus), Guangzhou 510275, China}

\begin{abstract}
Quantum metrology exploits quantum resources and strategies to improve measurement precision of unknown parameters.
One crucial issue is how to prepare a quantum entangled state suitable for high-precision measurement beyond the standard quantum limit.
Here, we propose a scheme to find optimal pulse sequence to accelerate the one-axis twisting dynamics for entanglement generation with the aid of deep reinforcement learning (DRL).
We consider the pulse train as a sequence of $\pi/2$ pulses along one axis or two orthogonal axes, and the operation is determined by maximizing the quantum Fisher information using DRL.
Within a limited evolution time, the ultimate precision bounds of the prepared entangled states follow the Heisenberg-limited scalings.
These states can also be used as the input states for Ramsey interferometry and the final measurement precisions still follow the Heisenberg-limited scalings.
While the pulse train along only one axis is more simple and efficient, the scheme using pulse sequence along two orthogonal axes show better robustness against atom number deviation.
Our protocol with DRL is efficient and easy to be implemented in state-of-the-art experiments.
\end{abstract}

\maketitle


\section{Introduction}\label{Sec1}

Quantum metrology studies how to exploit quantum resources and strategies to improve the estimation precision of unknown parameters~\cite{Giovannetti2006,Giovannetti2011}.
Generally, the information of an unknown parameter is encoded into a phase which can be precisely estimated via interferometric techniques in experiments~\cite{Gross2010,Lucke2011,Pezze2018}.
For interferometry with individual atoms, the sensitivity of the estimated phase can reach the so-called standard quantum limit (SQL)~\cite{Caves1981}, i.e., $\Delta\phi = \mathcal{O}(N^{-1/2})$ with $N$ the atom number.
However, this bound is not fundamental and can be surpassed by using multi-particle entanglement~\cite{Bengtsson2006,Lee2006,Esteve2008,Geza2014}.
Recent developments in quantum metrology focus on how to generate metrologically useful quantum entangled states and utilize them for phase estimation.

One kind of representative entangled quantum states that can provide sub-SQL phase sensitivity is spin-squeezed state~\cite{Wineland1992}.
Spin squeezed states can be prepared through the one-axis twisting (OAT) interaction, which is widely realized by light-mediated interactions~\cite{Leroux2010, Schleier-Smith2010, Zhang2015, Braverman2019} or atom-atom interaction within Bose condensed atoms~\cite{Kitagawa1993, Anders2001, Ae2002, Gross2010,Riedel2010} and the phase sensitivity can be scaled as $\Delta\phi = \mathcal{O}(N^{-2/3})$~\cite{Holland1993, Pezze2018}.
Apart from OAT, spin squeezing can be generated by two-axis counter-twisting (TACT) interaction, and the phase sensitivity can be improved to the Heisenberg limit, $\Delta\phi = \mathcal{O}(N^{-1})$.
However, this kind of spin squeezing is challenging to realize in experiments.
In addition to spin squeezed states, non-Gaussian entangled states such as twin Fock state and spin cat state are also promising candidates for achieving Heisenberg-limited phase sensitivity~\cite{Yurke1986, Holland1993,Giovannetti2006}.

The main obstacle against the applications of quantum entangled states in practice is the entanglement generation in realistic experiments.
Several theoretical schemes for preparing quantum entangled states such as adiabatic sweeping~\cite{Lee2006, Huang2015, Huang2018, Zhuang2020}, shortcut to adiabaticity~\cite{Julia2012,Lapert2012,Yuste2013} and optimal controls~\cite{Huang2018,Grond2009,Pichler2016, Sorelli2019} are developed.
However, the schemes are either time-consuming or too complicated to be implemented, which are hard to realize in state-of-the-art experiments.
Hence, developing fast and effective approaches for creating quantum entanglement is of great importance.

One promising way is to make use of machine learning, which has already attracted much attention~\cite{Carleo2019}.
In particular, deep reinforcement learning (DRL)~\cite{Sutton2018, Carleo2019} which can provide optimal decision strategies or policies based upon a well-defined target, is gradually applied in quantum physics~\cite{Dunjko2018,Carleo2019,Pantita2017,Fosel2018,Wallnofer2020,Chen2020,schafer2020,Saito2020,Rose2021}.
It can provide a machine learning (ML) model, often neural networks that is capable of optimizing a certain objective function by providing a well-designed time sequence of control procedures.
It is particularly suitable for seeking the optimal preparation of desired quantum states~\cite{Han2019,Chen2019,Predko2020,Haug2020,Han2020,Schuff2020,An2021,Fiderer2021,Guo2021}.
Recently, it is proposed that extreme spin squeezing can be achieved with OAT interaction using a sequence of rotation pulses designed via DRL~\cite{Chen2019}.
Although spin squeezing is a good metrological quantum resource, the most metrologically useful one is usually characterized by the quantum Fisher information (QFI) $F_Q$~\cite{Braunstein1994, Braunstein1996}.
Can we find out an experimentally feasible scheme to prepare the optimal quantum entangled state that maximizing $F_Q$ via DRL?
Can the prepared quantum entangled state suitable for practical quantum phase estimation?

In this work, we propose a scheme for preparing metrological useful entangled states based on OAT interaction with a sequence of rotation pulses designed via DRL.
In our scheme, the OAT interaction which is the key for entanglement generation, exists persistently during the state preparation.
Our scheme is inspired by the so-called twist and turn dynamics~\cite{Muessel2015, Sorelli2019} that is capable of generating spin squeezing efficiently.
In order to prepare the optimal quantum entangled state within a limited time $T$, a train of $\pi/2$ pulses is sophisticatedly applied~\cite{Chen2019}.
The time sequence of pulse train is obtained by maximizing $F_Q$ with the aid of DRL.

When considering $\pi/2$ pulses only along one axis, we find that only a few number of pulses can drive to a highly entangled state which enables the Heisenberg-limited scaling.
However, this protocol is sensitive to the atom number of the initial state.
In experiment, the atom number may not be well-defined and there will be a deviation from the atom number used in the DRL algorithm for designing the pulses.
This kind of atom number deviation may deviate the prepared state from the optimal one, hence degrade the ultimate measurement precision scaling.
To strengthen the robustness, we consider $\pi/2$ pulses along two orthogonal axes.
We find that although more pulses are required, it is more robust against atom number deviation.
To validate our scheme for phase estimation, we use the entangled states obtained by DRL as the input state to perform the Ramsey interferometry.
The associated phase measurement precision $\Delta \phi$ can still display the Heisenberg-limited scaling.
Besides, the scheme with $\pi/2$ pulses along two axes can also provide better robustness against the atom number deviation.
Our scheme via DRL provides a straightforward way to efficiently prepare optimal entangled states for quantum metrology,
and its robustness against the atom number deviation makes it feasible in realistic experiments.


\begin{figure*}[t]
  \includegraphics[width = 2.06\columnwidth]{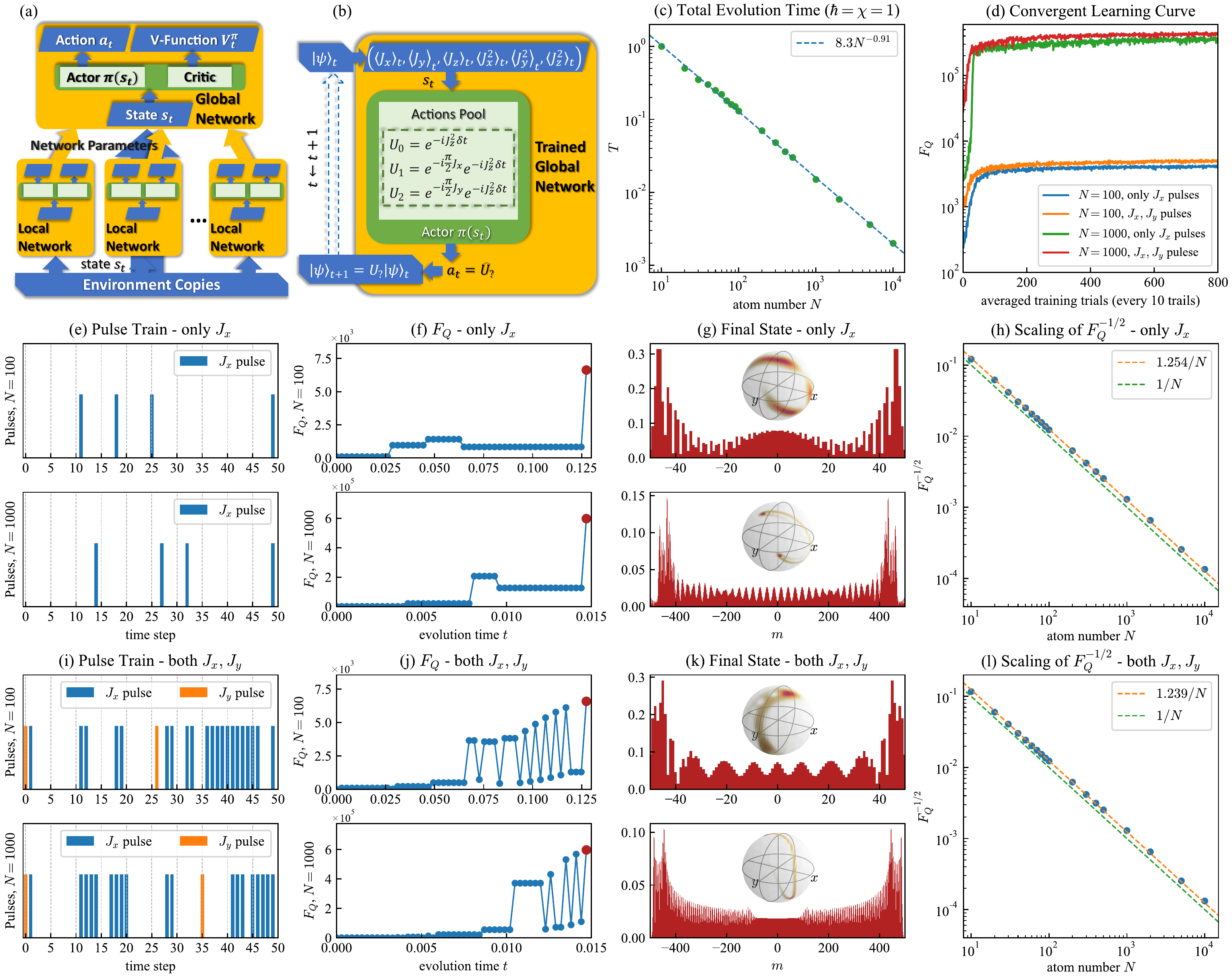}
  \caption{\label{fig:FIG1}
  (a) The sketch of Asynchronous Advantage Actor-Critic (A3C) algorithm, featuring local networks design and asynchronous updating of network parameters.
  (b) The sketch of the quantum state preparation process guided by A3C algorithm.
  In the $t$-th step, the trained network receives current state $s_t$ and then provides a certain action $a_t$, representing an operator $U_t$ participating in the next step.
  (c) The total evolution time $T$ versus atom number $N$ in our numerical calculations.
  A fitting function (the blue dashed line) is added, roughly showing an exponential relationship between $T$ and $N$.
  (d) The learning curves for $N=100$ and $1000$, including the results using only-$J_x$ and both-$J_x, J_y$ schemes.
  The convergent behaviors suggest the effectiveness of the whole learning process.
  The second row displays the results obtained by using only-$J_x$ scheme, i.e. the actions pool in (b) only contains $U_0$ and $U_1$.
  (e) The optimized pulse trains. Blue histograms are placed at the time step when a $\pi/2$ pulse along $x$ axis is applied.
  (f) The evolution of $F_Q$ during the state preparation process.
  (g) The optimized states with maximized $F_Q$.
  The corresponding Husimi functions on Bloch spheres are shown in the insets.
  (h) The scaling of $F_Q^{-1}$ versus the atom number $N$.
  Here, we fit the points by least square method and the fitting line is denoted by an orange line.
  The green dash line stands for the exact Heisenberg limit.
  The bottom row displays the results obtained by using both-$J_x, J_y$ schemes, i.e. the actions pool in (b) contains $U_0$, $U_1$ and $U_2$.
  (i) The optimized pulse trains. Blue and orange histograms represent the $\pi/2$ pulses along $x$ and $y$ axis, respectively.
  (j) The evolution of $F_Q$ during the state preparation process.
  (k) The optimized states.
  (l) The scaling of $F_Q^{-1}$ versus the atom number $N$.
  }
\end{figure*}

\begin{figure}[t]
  \includegraphics[width = \columnwidth]{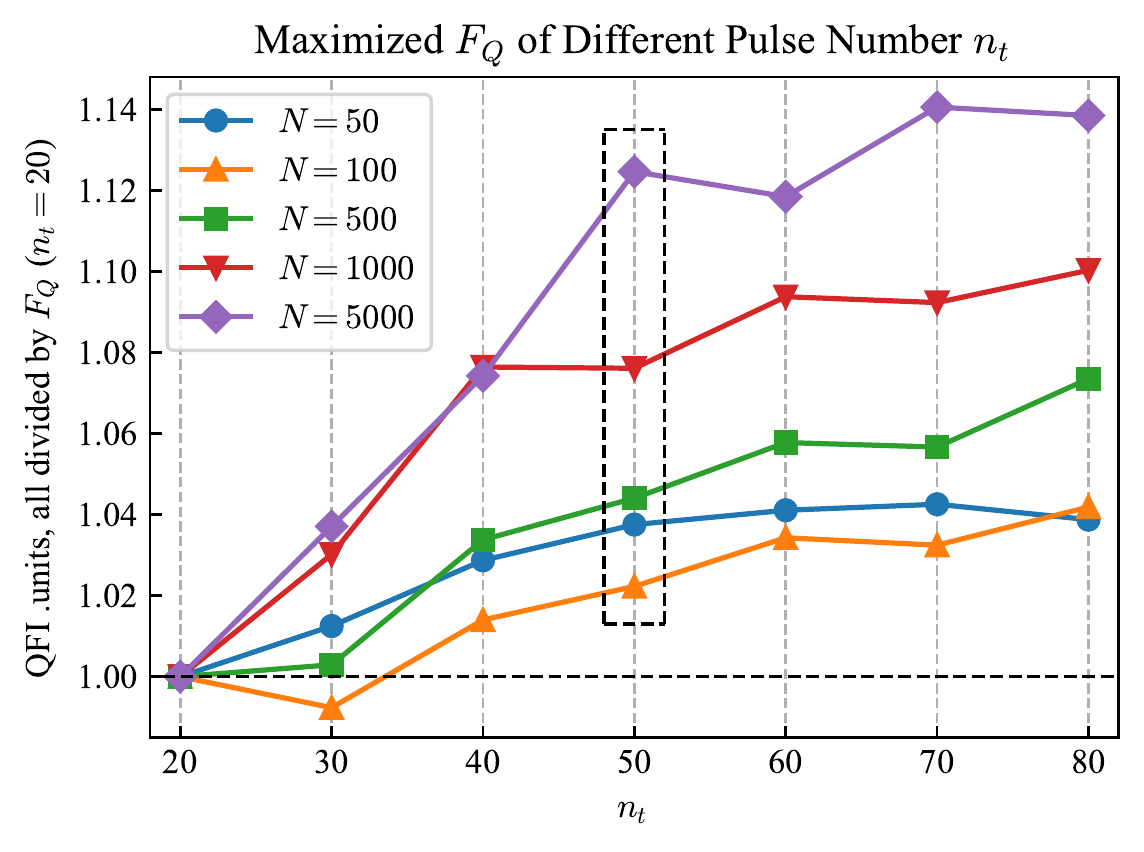}
  \caption{\label{fig:FIG2}
  The influence of the interval number $n_t$ in our DRL algorithm.
  We plot $F_Q$ of the final states $|\psi\rangle_T$ prepared by different pulse sequences trained by the same DRL algorithm with different pulse numbers $n_t$.
  The results of different atom number $N$ are shown, in which the quantities of $F_Q$ is divided by the minimum of those results with the same $N$.
  It is suggested that $n_t=50$ is a balanced choice with relatively large $F_Q$ and small amount of $n_t$ (surrounded by a black dash rectangle).
  }
\end{figure}

\begin{figure*}[t]
  \includegraphics[width = 2.06\columnwidth]{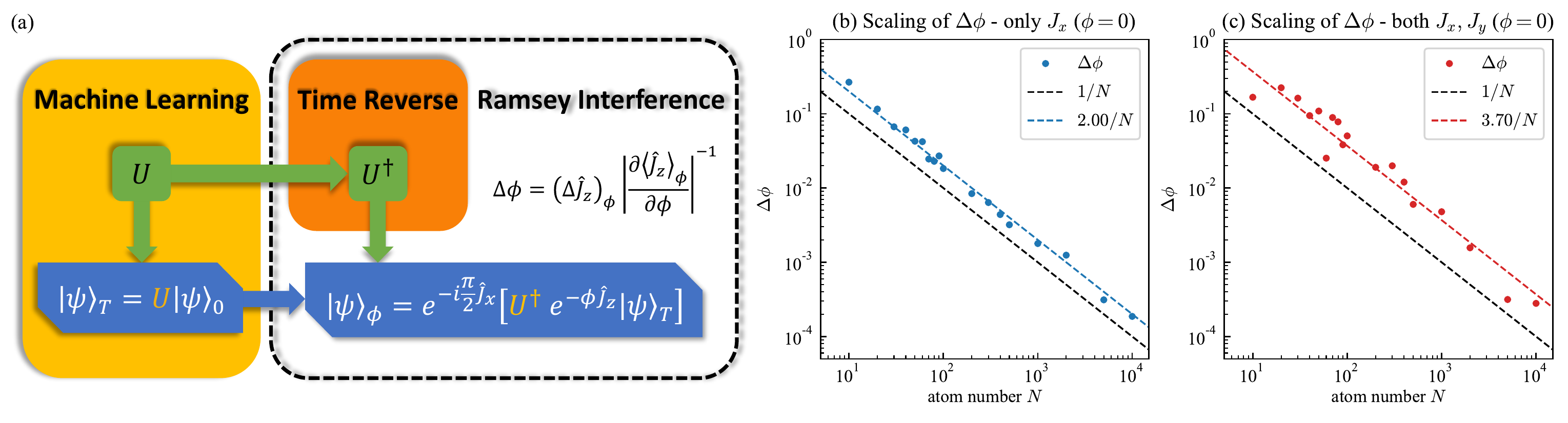}
  \caption{\label{fig:FIG3}
  (a) The sketch of Ramsey interferometry with time-reversal operations for phase estimation.
    An entangled state $|\psi\rangle_T$ is produced by the operation of $U$, which is obtained by our DRL algorithm.
    Then, the state $|\psi\rangle_T$ is input for the Ramsey interferometry, where a time-reversed operation $U^{\dagger}$ is used after the phase accumulation.
    Finally, applying a $\pi/2$ pulse and measuring the half relative population $\hat J_z$, one can extract the information of the estimated phase $\phi$.
    Here, we consider the phase is in the vicinity of $\phi=0$.
    The measurement precision scaling of estimated phase versus atom number $N$ obtained by (b) only-$J_x$ scheme and (c) both-$J_x,J_y$ scheme, respectively.
    %
    %
    The black dashed lines represent the Heisenberg limit $N^{-1}$, while the colored dashed lines are the Heisenberg-limited scaling obtained by fitting the numerical results.
  }
\end{figure*}


\section{Entanglement generation via Deep Reinforcement Learning}\label{Sec2}

\subsection{Preparation of quantum entangled state}
We consider an ensemble of $N$ two-level identical atoms whose Hamiltonian ($\hbar=1$) is given by
$H = \chi \hat J_z^2 + \Omega \hat J_\gamma + \delta \hat J_z$.
Here, $\hat J_\gamma = \hat J_x \cos{\gamma} + \hat J_y \sin{\gamma}$ and $\hat J_{\alpha} = \sum_{l} \sigma_{\alpha}^{(l)}/2$ ($\alpha=x,y,z$) are the collective spin operators with the Pauli matrices $\sigma_{\alpha}^{(l)}$ for the $l$-th atom~\cite{Gross2010}.
The system state can be expanded in the Dicke basis $\hat J_z |m\rangle = m|m\rangle$ with $m=-N/2, -N/2+1, ..., N/2$.
The Hamiltonian contains three terms.
The first term $\chi \hat J_{z}^{2}$ denotes the atom-atom interaction, which is the key for realizing one-axis twisting (OAT) dynamics~\cite{Riedel2010,Gross2010}.
The second term $\Omega \hat J_\gamma$ is the coupling between the two atomic levels.
The third term $\delta \hat J_z$ is the bias or detuning.
The Hamiltonian $H$ can be applied to Bose condensed atoms occupying two hyperfine states~\cite{Hall1998, Cirac1998} or a single-component condensate trapped in a double-well potential~\cite{Schumm2005, Hall2007, Jo2007}.
The parameters $\chi$, $\Omega$ and $\delta$ can be well controlled via external fields in experiments~\cite{Gross2012,Pezze2018}.

The first and significant step for quantum metrology is the entangled state preparation.
Initially, the system state is usually prepared in a coherent spin state (CSS)~\cite{Radcliffe1971,Arecchi1972}
\begin{equation}\label{initial_state}
  |\psi\rangle_{0} = e^{-i\frac{\pi}{2}\hat J_y}|\pi, 0\rangle_{\mathrm{CSS}},
\end{equation}
which is rotated by a $\pi/2$ pulse along the $y$ axis~\cite{Zhang1990,Pezze2018} from the state $|\pi, 0\rangle_{\mathrm{CSS}} = |\uparrow\rangle^{\otimes N}$ with all $N$ atoms in $|\uparrow\rangle$.
The OAT dynamics can squeeze the CSS to a spin squeezed state.
There exists an optimal evolution time $T_{os}$ that extreme spin squeezing can be achieved~\cite{Chen2019}.
Apart from spin squeezing, the metrological ability of a quantum state can also be characterized by QFI.
Generally, maximizing $F_Q$ can obtain the optimal input state for attaining the best precision bounds~\cite{Paris2009,Giovannetti2011,Geza2014}.
Thus, we use QFI as a metric to find out the optimal input state for phase estimation below.
For an input state $|\psi\rangle$, the QFI for phase estimation can be defined as~\cite{Pezze2018}
\begin{equation}\label{qfi}
  F_Q = 4[_t\langle\psi'(\theta)|\psi'(\theta)\rangle_t -
    |_t\langle\psi'(\theta)|\psi(\theta)\rangle_t|^2],
\end{equation}
where $|\psi(\theta)\rangle_t = e^{-i\theta\hat J_z}|\psi\rangle$ and
$|\psi'(\theta)\rangle_t = -i \hat J_z|\psi(\theta)\rangle_t$.
Therefore, the ultimate precision bound can be given by $F_Q^{-1/2}$~\cite{Braunstein1994, Braunstein1996}.
To speed up the entanglement generation, in the stage of state preparation, we apply some pulses and therefore the system obeys
\begin{equation}\label{bjj_in}
  H = \chi \hat J_z^2 + \Omega_x(t)\hat J_x + \Omega_y(t)\hat J_y,
\end{equation}
where $\Omega_x(t)$ and $\Omega_y(t)$ are time-dependent functions describing the applied pulses.
Consider the total evolution time $T$ is around $T_{os}$, and we divide $T$ equally into $n_t$ intervals and each interval length $\delta t = T/n_t$.
At each interval, one can choose to apply a $\pi/2$ pulse along $x$ or $y$ axis with $\int_t^{t+\delta t}{\Omega_{x,y}(t')dt'} = \pi/2$, or turn off the coupling $\Omega_{x,y}=0$ to let the state evolve barely under OAT interaction.

Our goal is to find the optimized pulse train to generate the input state $|\psi\rangle$ that maximizing $F_Q$ within $T$ from the initial CSS $|\psi\rangle_{0}$.
To accomplish this goal, we adopt the technique from machine learning (ML).
The optimization process will be guided by an ML model obtained from DRL.
In the following, we will introduce the DRL algorithm and show the optimization results in detail.

\subsection{DRL algorithm}
To obtain the optimal control, the optimization process will be guided by a DRL algorithm.
Briefly, the DRL algorithm requests certain information about the current state $|\psi\rangle_t$ for the $t$-th time step ($t\in[1, n_t]$), and determines the evolution happening in the next $(t+1)$-th time step with an optimal policy.
As one of the DRL algorithms, here we adopt the so-called Asynchronous Advantage Actor-Critic (A3C) algorithm~\cite{Mnih2016} to accomplish our goal.
It is based on a common actor-critic algorithm while designed in an asynchronous structure, as sketched in Fig.~\ref{fig:FIG1} (a).
Generally it uses neural networks to find an appropriate decision.
The network parameters are updated via adaptive momentum gradient decent method (ADAM)~\cite{Paszke2017}.
The asynchronous structure of A3C is beneficial for the stability of the learning process and makes it fast to converge.
The learning process also becomes more efficient because the local network design is naturally parallel processing which can take full advantages of the multiple process units in the computing hardware.

Next, we show how to find the optimized pulse train in the framework of DRL algorithm.
%
As shown in Fig.~\ref{fig:FIG1} (b), at every time step $t$ the algorithmic state $s_t$ needed to know and feed into the algorithm is some expectation values of the evolved quantum state $|\psi\rangle_t$. $s_t$ can be encoded in a tuple with the following six expectation:
$(\langle \hat J_{x} \rangle_t, \langle \hat J_{x} \rangle_t, \langle \hat J_{x} \rangle_t,
\langle \hat J_{x}^{2} \rangle_t, \langle \hat J_{x}^{2} \rangle_t, \langle \hat J_{x}^{2} \rangle_t)$.
It should be mention that, these six expectation quantities are the intermediate variables in the algorithm.
They are only calculated numerically~\cite{Chen2019} and do not need to be measured in experiments.
%
%
%
Then the action $a_t$ is obtained after receiving $s_t$, which is an evolution operator $U_t$ chosen from the action pool containing three candidates:
\begin{equation}\label{actions}
\begin{split}
  \hat U_0&=e^{-i\chi\hat J_z^2\delta t}, \\
  \hat U_1&=e^{-i\frac{\pi}{2}\hat J_x}e^{-i\chi\hat J_z^2\delta t}, \\
  \hat U_2&=e^{-i\frac{\pi}{2}\hat J_y}e^{-i\chi\hat J_z^2\delta t}.
\end{split}
\end{equation}
Finally a reward $r_t$ related to the QFI of evolved state $F_Q^{(t)}$ is calculated. The reward will be described later.
%
%

In this work, we consider two schemes, ``only-$J_x$" and ``both-$J_x,J_y$".
The former one only using $\pi/2$ pulses along $x$ axis, in which $U_t$ is chosen only from $U_0$ and $U_1$.
While for the latter one, $\pi/2$ pulses along $x$ and $y$ axis are both considered, i.e., $U_t \in \{\hat U_0, \hat U_1, \hat U_2\}$.
Then, the unitary evolution $|\psi\rangle_{t+1} = U_t|\psi\rangle_{t}$ is performed, and the consequent state $|\psi\rangle_{t+1}$ will participate the evolution at the next time step $t+1$ sequentially.
Thus, the final prepared state can be written as
\begin{equation}\label{entagle}
  |\psi\rangle_T = U |\psi\rangle_0 = \prod_{t=1}^{n_t} U_t |\psi\rangle_0,
\end{equation}
where the initial state is given by Eq.~\eqref{initial_state}.
To maximize $F_Q$ of $|\psi\rangle_{T}$, in each step we numerically calculate the QFI $F_Q^{(t)}$ for $|\psi\rangle_t$  to obtain the reward $r_t$ of the $t$-th step.
The calculation of the total reward $R_{\mathrm{tot}}$ is then made after $n_t$ evolution steps.
Finally, a specific pulse sequence $(U_1, U_2, ..., U_{n_t})$ can be generated from the optimal policy within the DRL algorithm.

The total reward $R_{\mathrm{tot}}$ is originally the accumulated reward of $n_t$ time steps as $R_{\mathrm{tot}} = \sum_{t=0}^{n_t}r_t$~\cite{Chen2019},
while in our DRL algorithm the $n_t$ rewards are requested all in once after total evolution time $T$,
by denoting the reward of the $t$-th step as the largest reward among the rest steps after time $t$, as:
\begin{equation}\label{reward}
  r_t = \max_{t<i<n_t}F_Q^{(i)}
\end{equation}
This non-step-wise design of reward allows us to denote every $r_t$ after knowing $F_Q^{(0\sim n_t)}$, which is beneficial for the training stability, efficiency and capability of convergent.
Another advantage of this definition~\eqref{reward} is that in each training epoch the DRL algorithm can somewhat comprehend that
the optimization task is fulfilled within $n_t$ steps so that the ML model can reach similar optimum once $n_t$ is large enough, see Fig.~\ref{fig:FIG2}.
In addition, we use two separated neural networks as actor and critic network.
The benefit of this separation is that different quantities of $F_Q$ from different atom numbers $N$ can be greatly balanced.
The parameters of our algorithm, including structure of the neural networks and the learning rate,
do not need to be adjusted in the face of different atom number situations and can achieve convergence at the same rate, see Fig.~\ref{fig:FIG1} (d).

\subsection{Results with DRL}
In our numerical simulations, we choose $\chi=1$ and $n_t=50$.
The total evolution time $T$ is chosen near the optimal squeezing time, which can be determined numerically.
The relation between $T$ and $N$ is shown in Fig.\ref{fig:FIG1}~(c), roughly an exponential dependence.
For example, for $N=100$ and $1000$ we have $T=0.13$ and $0.015$, respectively.
Starting from an initial $|\psi\rangle_0$ with a fixed $N$, we can obtain the maximized $F_Q$ and the corresponding prepared quantum state $|\psi\rangle_{T}$ with the help of DRL.
Here, we display results of two representative cases ($N = 100$ and $1000$) using only-$J_x$ scheme and both-$J_x, J_y$ scheme, see Fig.\ref{fig:FIG1}~(e$\sim$h) and (i$\sim$l), respectively.
In Fig.~\ref{fig:FIG1}~(d), the learning curves of DRL for both schemes with $N=100$ and $1000$ are given.
It is shown that, after $8000$ trails of learning the $F_Q$ of the final states $|\psi\rangle_{T}$ are optimized and converge to saturated values, indicating a successful optimization.

The associated pulse trains optimized by our DRL algorithm for $N=100$ and $1000$ are shown as histograms in Fig.~\ref{fig:FIG1}~(e) for only-$J_x$ scheme and in Fig.~\ref{fig:FIG1}~(i) for both-$J_x,J_y$ scheme, where blue and orange histograms stand for $\pi/2$ pulses along $x$ and $y$ axis, respectively.
The corresponding time-evolutions of the $F_Q$ are shown in Fig.~\ref{fig:FIG1}~(f) and (j).
The $F_Q$ of the optimal prepared states $|\psi\rangle_{T}$ are highlighted by red dots, and the associated distributions of $|\psi\rangle_{T}$ are shown in in Fig.~\ref{fig:FIG1}~(g) and (k).

The optimized $F_Q$ of the prepared states using only-$J_x$ scheme and both-$J_x,J_y$ scheme are nearly the same, with the latter mostly being a little larger than the former.
The final prepared states $|\psi\rangle_{T}$ become non-Gaussian with two humps appear near $|m=\pm N/2\rangle$, see the Husimi distribution on the generalized Bloch sphere and the probability distribution.
However, the probability distribution of $|\psi\rangle_{T}$ using both-$J_x,J_y$ scheme is more rugged than the one using  only-$J_x$ scheme.
Essentially, we find that the scaling of $F_Q$ versus $N$ of the two schemes can both approach the Heisenberg limit. Here, we use least square method to fit the results and the fitting formula are displayed in the legends.
Similarly, the both-$J_x,J_y$ scheme outperforms the only-$J_x$ scheme with a slightly smaller constant.
It is evident that the method with DRL algorithm is promising for developing Heisenberg-limited metrology protocols.

On the other hand, the optimized pulses trains for these two schemes are much different.
We can see that, for both $N=100$ and $1000$, only four $\pi/2$ pulses along $x$ axis is needed.
With a final pulse applying at the final time step, the state can abruptly evolve to the optimal one.
The corresponding $F_Q$ suddenly jump to a large value.
While for both-$J_x,J_y$ scheme, more $\pi/2$ pulses along $x$ axis with few $\pi/2$ pulses along $y$ axis are needed.
Thus, the pulse trains for only-$J_x$ scheme is much sparse and simple, which will be more feasible in realistic experiments.
For a fixed $N$, whatever by using only-$J_x$ scheme or both-$J_x,J_y$ scheme, we can find the optimal control for preparing the optimal state within $T$ with the help of DRL algorithm.
However, the optimized pulse trains are always discrepant with different $N$ and $T$.
Thus, we need to know the atom number $N$ roughly in advance to design the corresponding optimal pulse sequence.

The interval number $n_t$ we divide the total evolution time $T$ may slightly influence the optimization results.
The resultant $F_Q$ of the final states $F_Q$ with different $n_t$ are shown in Fig.\ref{fig:FIG2}.
It is shown that more pulses enable to push the optimization even better but the growth decreases when $n_t>50$, especially for large $N$.
Thus, we find that $n_t=50$ is a balanced choice in condition that the structure of the two networks and hyperparameters in our DRL algorithm also remain unchanged.
Despite that with increasing $n_t$ the $F_Q$ of the prepared state may be slightly larger, it requires more carefully designed algorithm parameters and increases operation complexity.


\begin{figure*}[t]
  \includegraphics[width = 2.06\columnwidth]{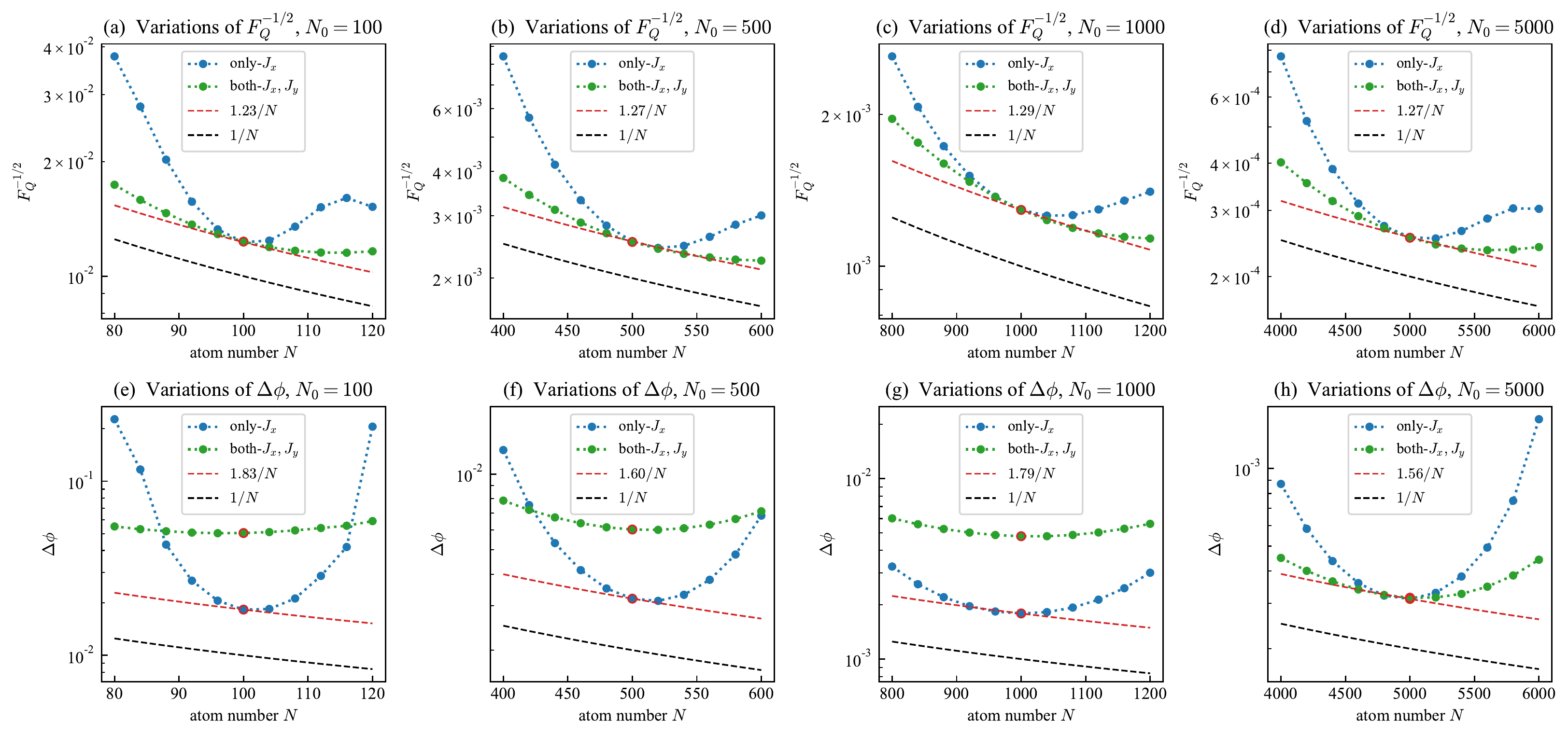}
  \caption{\label{fig:FIG4}
  The robustness against atom number deviation.
  Here, the pulse trains are obtained by DRL with (a) $N=100$, (b) $N=500$, (c) $N=1000$,  and (d) $N=5000$, respectively.
  The ultimate precision bounds are obtained by the same pulse train in condition of different atom number $N$.
  The blue points connected by blue dashed lines are results obtained by only-$J_x$ scheme, and green points connected by green dashed lines are results obtained by both-$J_x,J_y$ scheme.
  The black dashed lines stands for the exact Heisenberg limit $1/N$,
  and the red dashed lines is the Heisenberg-limited scaling $\propto 1/N$ passing through the point of the $F_Q^{-1/2}$ of original values of $N$ that is highlighted by red circles.
  (e)-(h) Phase measurement precision $\Delta \phi$ with time-reversal Ramsey interferometry for the same situations and plotted in the same manner with (a)-(d), respectively.
  Despite the absolute value using both-$J_x,J_y$ scheme is mostly a little worse, the both-$J_x,J_y$ scheme displays better robustness against deviation of atom number $N$.
  }
\end{figure*}

\section{Phase estimation via time-reversal Ramsey interferometry}\label{Sec3}

Generally, QFI only sets the ultimate measurement precision bound, but it may not always be attained.
To validate metrological usefulness of the prepared states via DRL, we implement the Ramsey interferometry for phase estimation~\cite{Ramsey1963,Gross2010,Lucke2011} by inputting the prepared states $|\psi\rangle_T$.

For a conventional Ramsey interferometry, the whole process consists of a phase accumulation sandwiched by two $\pi/2$ pulses~\cite{Geza2014,Rafal2015}.
Since we start from an initial CSS, it is suitable to use time-reversal protocol.
Here, we consider a time-reversal protocol: a disentangling operation $U^{\dagger}$ after the phase accumulation process~\cite{Frowis2016, Davis2016}, which is implemented by a reverse of $U$ in Eq.~\eqref{entagle}.
As sketched in Fig.~\ref{fig:FIG3}~(a), the final state after Ramsey interferometry is thus:
\begin{equation}
  |\psi\rangle_\phi = e^{-i\frac{\pi}{2}\hat J_x} U^{\dagger} e^{-\phi \hat J_z} |\psi\rangle_T.
\end{equation}
The time-reversal operation can be achieved by changing the sign of the entangling Hamiltonian~\cite{Davis2016}.
This can be realized in various synthetic quantum systems, such as atom-cavity system~\cite{Colombo2021} and cold atom system~\cite{Linnemann2016}.

Then the measurement precision of $\phi$ can be calculated by using error propagation formula~\cite{Gross2012}:
\begin{equation}\label{Delta_phi}
  \Delta\phi = \frac{(\Delta \hat J_z)_\phi}{|\partial\langle \hat J_z \rangle_\phi / \partial\phi|},
\end{equation}
where $(\Delta \hat J_z)_\phi = \sqrt{\langle \hat J_z^2 \rangle_\phi - \langle \hat J_z \rangle_\phi^2}$, the subscript $\phi$ indicates the expectation with respect to $|\psi\rangle_\phi$.
Here, we consider the estimated phase is tiny which is in the vicinity of $\phi=0$.

The corresponding scalings of measurement precision versus $N$ are shown in Fig.~\ref{fig:FIG3}~(b) and (c).
The resultant phase measurement precisions are given as blue (only-$J_x$ scheme) and red points (both-$J_x,J_y$ scheme), respectively.
Despite the scaling is a bit deviated from the ultimate bounds of $F_Q$ in Fig.~\ref{fig:FIG1}~(h) and (l), the estimated phase measurement precision for only-$J_x$ and both-$J_x,J_y$ schemes still show Heisenberg-limited scaling as expected.
This suggests the optimized entangled state we prepare by using DRL algorithm also has great potential for Heisenberg-limited phase estimation with Ramsey interferometry.

The only-$J_x$ scheme shows a smoother scaling and closer to the Heisenberg limit, $2.0 / N$ compared to $3.7 / N$ that obtained by both-$J_x,J_y$ scheme.
This may result from the addition of $U_2$ pulses in Eq.~\eqref{actions}, while in the next section we will see that the participation of $U_2$ can contributes to a better robustness against the deviation of atom number $N$.


\section{Robustness against atom number deviation}\label{Sec4}
Finally, we discuss the robustness of our schemes against the atom number deviation.
As it is mentioned in Sec.~\ref{Sec2}, the optimal pulse sequence obtained by DRL depends on the atom number $N$ and total evolution time $T$.
In our numerical calculations, the initial state $|\psi\rangle_0$ is assumed to be a pure state with a well-defined atom number $N$.
In practice, $T$ can be precisely controlled but the estimation of atom number $N$ may be inaccurate.
The atom number in experiment may not be the same as expected.
There may be a deviation between the atom number in experiment and the one set in the DRL algorithm for designing the pulses.
%
%
%
Therefore, it is necessary to figure out the robustness of our scheme when this kind of atom number deviation exists.

We perform the robustness evaluation by applying the optimized pulse train of atom number $N$ to the situation with other atom number in the range of $[0.8 N, 1.2 N]$, i.e., the deviation of atom number is assumed up to $\pm 20\%$.
%
%
The results with $N=100, 500, 1000$ and $5000$ are shown in Fig.~\ref{fig:FIG4}, including the $F_Q$ and phase measurement precision $\Delta \phi$ via time-reversal Ramsey interferometry, using only-$J_x$ scheme and both-$J_x,J_y$ scheme.
The red dashed lines are added for reference, representing the Heisenberg-limited scalings passing the results of only-$J_x$ scheme cases without deviations.
Ideally, the results should be close to the red dashed lines.

It turns out that the deviation of $N$ makes the resultant ultimate precision bound $F_Q^{-1/2}$ degraded, and the results of $\Delta \phi$ also become worse.
Compared with only-$J_x$ scheme, both-$J_x,J_y$ scheme show better robustness against atom number deviation.
As it is shown in Fig.~\ref{fig:FIG4}~(a)-(d), the $F_Q$ keeps in the same level when there is no deviation of $N$,
and the $F_Q$ using both-$J_x,J_y$ decrease much less than those using only-$J_x$ scheme.
The cases of $\Delta \phi$ is shown in Fig.~\ref{fig:FIG4}~(e)-(h), showing the same manner of degradation with these two schemes.
Although the phase measurement precision using both-$J_x,J_y$ scheme is worse than those using only-$J_x$ scheme for most $N$ as shown in FIG.~\ref{fig:FIG3} (b, c), the robustness of the former scheme is better than the latter.

It suggests that the pulse trains optimized by the DRL algorithm is practicable even though the atom number $N$ of the system cannot be estimated accurately.
If the atom number deviation is small in experiment, one may give priority to use the only-$J_x$ scheme for phase estimation.
Otherwise, the both-$J_x,J_y$ scheme which can show better robustness against atom number deviation, may become favorable.


\section{Conclusion and Discussion}\label{Sec5}

We have presented an efficient and robust scheme for preparing entangled state with DRL algorithm and demonstrated their metrological usefulness with the Ramsey interferometry for phase estimation.
We implemented the quantum state preparation through only-$J_x$ scheme or both-$J_x,J_y$ scheme, referring to the OAT dynamics with pulse sequence along only one axis or along two orthogonal axes, respectively.
The system starts from a CSS, then reaches an optimal entangled state under a pulse train optimized by DRL.
The quantum state preparation process is accomplished within a short time duration and the ultimate precision bounds exhibit the Heisenberg-limited scaling.
Further, the Heisenberg-limited scaling can be maintained by performing the Ramsey interferometry, which verify the usefulness of our schemes in experiments.
We use the A3C algorithm~\cite{Mnih2016} whose actor and critic networks are separately established.
It makes our algorithm equally effective and efficient for different atom number cases from $N=10$ to $10000$ without reforming the neural networks and parameters of the DRL algorithm.
Besides a non-step-wise reward design makes the training process feasible and stable, similarly successful when the total number of pulses $n_t$ is sufficient.

The only-$J_x$ scheme and both-$J_x,J_y$ scheme have different advantages.
On one hand, the pulse trains of only-$J_x$ scheme provided by DRL algorithm is much more simple, and the scaling of phase measurement precision is better than that of both-$J_x,J_y$ scheme.
On the other hand, we find that the entangled states prepared by both-$J_x,J_y$ scheme have better robustness against atom number deviation.
Therefore only-$J_x$ scheme can be used when one wants to simplify the process of state preparation and the deviation of atom number can be well controlled, while the both-$J_x,J_y$ scheme is considerable when the robustness against atom number deviation matters more.

Our algorithm can be used as an offline optimization for quantum entangled state preparation in synthetic many-body quantum systems, such as cold atoms~\cite{Gross2010, Riedel2010}, and trapped ions~\cite{Kevin2021}.
Online optimization is also feasible when the QFI is extractable~\cite{Strobel2014} while accompanying a huge consumption of time, which might be solved by starting from results provided by sufficient offline optimizations.
In the future, the effects of decoherence and imperfect pulse shape can also be taken into account, which will be more feasible for practical experiments.

\acknowledgments{
  This work is supported by the National Natural Science Foundation of China (12025509, 11874434),
  the Key-Area Research and Development Program of GuangDong Province (2019B030330001),
  and the Science and Technology Program of Guangzhou (201904020024). M.~Z. is partially supported by the National Natural Science Foundation of China (12047563).
  J.~H. is partially supported by the Guangzhou Science and Technology Projects (202002030459).
  }

\bibliography{DRL_QFIoptimize}

\end{document}